\documentclass{natureprintstyle}
\usepackage{graphicx}
\usepackage{amssymb,amsmath}
\usepackage[pdftex,breaklinks]{hyperref}
\hypersetup{
    colorlinks=true,
    linkcolor=blue,
    citecolor=blue,
    urlcolor=blue
}
\date{August 7, 2019 \\ \\
\sf{authors' version of \emph{Nature Astronomy} invited review
article \\
final version available at
\url{http://dx.doi.org/10.1038/s41550-019-0858-0}}}
\newcommand{\kms}{km~s$^{-1}$}
\newcommand{\sun}{\odot}
\newcommand{\Msun}{M$_\sun$}

\newcommand\snia{SN~Ia}

\newcommand\sneia{SN~Ia} % SWJ prefers not to use SNe -- too ugly!
\newcommand\sneIa{\sneia} % SWJ prefers not to use SNe -- too ugly!
\newcommand\NaiD{Na I D}

\newcommand\Caii{Ca II}
\newcommand\MCh{\emph{M}$_{\text{Ch}}$}
\title{Observational Properties of Thermonuclear Supernovae}

\author{Saurabh~W.~Jha$^{1,2}$, Kate~Maguire$^{3,4}$, Mark~Sullivan$^{5}$}

\begin{document}

\maketitle

\begin{affiliations}
\item Department of Physics and Astronomy, Rutgers, the State University of
New Jersey, Piscataway NJ, USA
\item Center for Computational Astrophysics, Flatiron Institute, New York,
NY, USA
\item Astrophysics Research Centre, School of Mathematics and Physics,
Queen's University Belfast, UK
\item School of Physics, Trinity College Dublin, Ireland
\item School of Physics and Astronomy, University of Southampton, Southampton,
SO17 1BJ, UK
\end{affiliations}

\begin{abstract}
\vskip 21pt
The explosive death of a star as a supernova is one of the most dramatic
events in the Universe. Supernovae have an outsized impact on many areas of
astrophysics: they are major contributors to the chemical enrichment of the
cosmos and significantly influence the formation of subsequent generations of
stars and the evolution of galaxies. Here we review the observational
properties of \emph{thermonuclear} supernovae, exploding white dwarf stars
resulting from the stellar evolution of low-mass stars in close binary
systems. The best known objects in this class are type Ia supernovae (\sneia),
astrophysically important in their application as standardisable candles to
measure cosmological distances and the primary source of iron group elements
in the Universe. Surprisingly, given their prominent role, \sneia\ progenitor
systems and explosion mechanisms are not fully understood; the observations we
describe here provide constraints on models, not always in consistent ways.
Recent advances in supernova discovery and follow-up have shown that the class
of thermonuclear supernovae includes more than just \sneia, and we
characterise that diversity in this review.
\end{abstract}

The modern classification scheme for supernovae traces back to
Minkowski\cite{1941PASP...53..224M} who in 1941 split ``Type I'' from ``Type
II'' supernovae based on optical spectra. Further subdivision of these basic
classes has continued on an empirical
basis\cite{1997ARA&A..35..309F,Gal-Yam2017}, and in our review we describe the
observational properties of what are now called \sneia, along with other
similar objects. The observational classification effort arises from a desire
for \emph{physical} understanding of these objects, explaining our use of the
term \emph{thermonuclear} supernovae in the title. That categorisation is
based on the explosion mechanism: objects where the energy released in the
explosion is primarily the result of thermonuclear fusion. Given our current
state of knowledge, we could equally well call this a review of the
observational properties of \emph{white dwarf} supernovae, a categorisation
based on the kind of object that explodes. This is contrasted with
\emph{core-collapse} or \emph{massive star} supernovae, respectively, in the
explosion mechanism or exploding object categorisations. Unlike those objects,
where clear observational evidence exists for massive star progenitors and
core-collapse (from both neutrino emission and remnant pulsars), the
\emph{direct} evidence for thermonuclear supernova explosions of white dwarfs
is limited\cite{2012ApJ...744L..17B,2014Natur.512..406C} and not necessarily
simply interpretable\cite{2013ApJ...769...67P,2014Sci...345.1162D}.
Nevertheless, the indirect evidence is strong, though many open questions
about the progenitor systems and explosion mechanisms remain.

\sneia\ are important both to the evolution of the Universe
and to our understanding of it. As standardisable candles whose distance can
be observationally inferred\cite{Phillips93}, \sneia\ have a starring role in
the discovery of the accelerating expansion of the
Universe\cite{riess98,perlmutter99} and in measurement of its current
expansion rate\cite{2016ApJ...826...56R}. \sneia\ are also major contributors
to the chemical enrichment of the Universe, producing most of its
iron\cite{Nomoto13} and elements nearby in the periodic table. Because of the
stellar evolutionary timescales involved, the enrichment of these elements
occurs differently from other elements whose main origin is in massive star
supernovae.

Here we review the observational properties of thermonuclear supernovae,
including both normal \sneia\ and related objects. We describe the photometric
and spectroscopic properties of \sneIa\ in section \ref{sec:snIa}, and their
environments and rates in section \ref{sec:environs}. Evidence has been growing
that not all thermonuclear explosions of white dwarfs result in ``normal''
\snia; we discuss related supernovae in section \ref{sec:zoo}. In this review
article we provide a broad overview supplemented by further discussion of the
newest developments. Our reference list is limited and thus necessarily
incomplete. We have chosen to highlight illustrative, recent works with a strong
bias towards observations rather than theory or models. These deficiencies are
rectified in recent reviews that cover many of these topics in more
detail\cite{Hillebrandt13,2017hsn..book.....A,2017suex.book.....B}.

\clearpage
\section{Type Ia Supernovae}
\label{sec:snIa}

\subsection{Energetics and light curve properties:} 
The runaway thermonuclear explosion of a carbon-oxygen white dwarf to
iron-group elements releases on the order of 10$^{51}$ erg as kinetic energy
that unbinds the star. The expanding ejecta travel at $\sim$10,000 \kms\ and
cool rapidly. The luminosity of \sneia\ is subsequently powered by the decay
of radioactive elements that were synthesised in the
explosion\cite{Pankey62,colgate}. The primary power source is the isotope
nickel-56, which decays to cobalt-56 with a half-life of 6.1 days, and which
in turn decays with a half-life of 77.3 days to stable iron-56. The peak
\snia\ bolometric luminosity is typically of the order of 10$^{43}$ erg
s$^{-1}$, with 0.3--0.8 \Msun\ of iron-56 ultimately produced in each event.
The majority ($\sim$85 \%) of the luminosity of a
\snia\ emerges at optical wavelengths and this is where they have been best
studied to date. \emph{Arnett's rule}\cite{arnett} says the peak luminosity of
the SN is proportional to the mass of nickel-56 produced in the explosion,
though in general this is only approximately true\cite{Katz13,Khatami19}.

\begin{figure*}
\begin{center}
\includegraphics[width=0.9\textwidth]{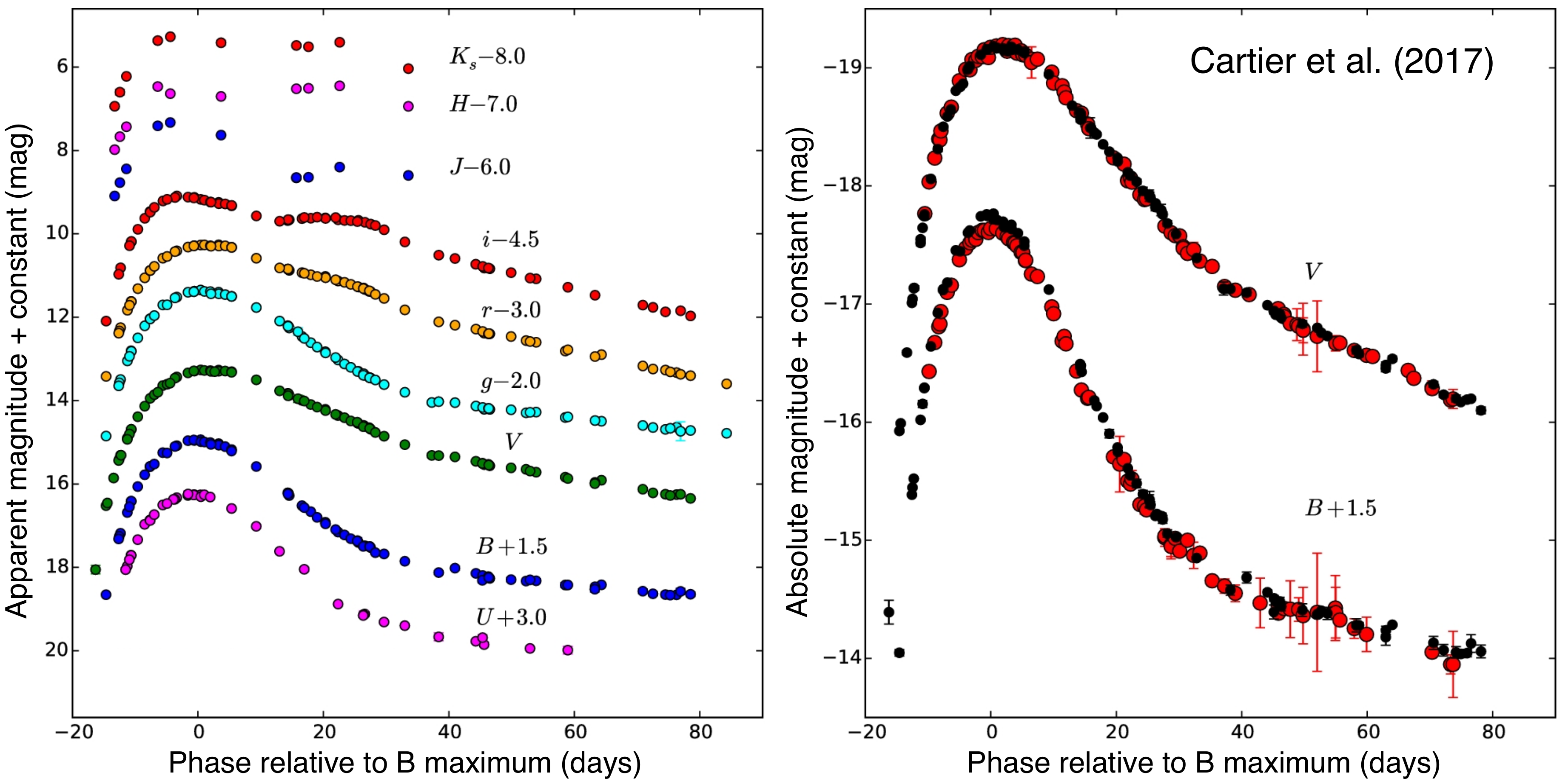}
\vskip -0.1in \caption{\textbf{\snia\ light curves.} 
\textit{Left panel:} Optical and near-infrared light curve of the type-Ia
SN~2015F. \textit{Right panel:} Comparison of the \textit{B} and \textit{V}
band light curves of SN~2015F (black points) and SN~2004eo (red points)
showing the similarity between these two \snia. The error bars displayed are
1-$\sigma$ uncertainties. This figure is adapted from
ref.~\citen{Cartier2017}.
 \label{fig:cartier}}
\end{center}
\end{figure*}

The optical light curves of normal \sneia\ are relatively homogeneous (and can
be standardised as discussed below), with a rise to peak luminosity in
$\sim$20 days, and a slow decline after peak before settling on to an
exponential decay phase after $\sim$50 days (Figure \ref{fig:cartier}). The
infrared light curves of \sneia\ are characterised by a secondary peak 20--30
d after maximum light that is not seen at bluer wavelengths, thought to be
driven by the recombination of doubly to singly ionised
iron\cite{Kasen2006,kromer--10}. \sneia\ peak in the ultraviolet (UV) roughly
15--20 d after explosion, slightly before the optical, but with a much lower
flux at most epochs ($<$10\% of the optical luminosity at peak) due to
strong iron-group line blanketing opacity.

\begin{figure*}
\begin{center}
\includegraphics[width=0.9\textwidth]{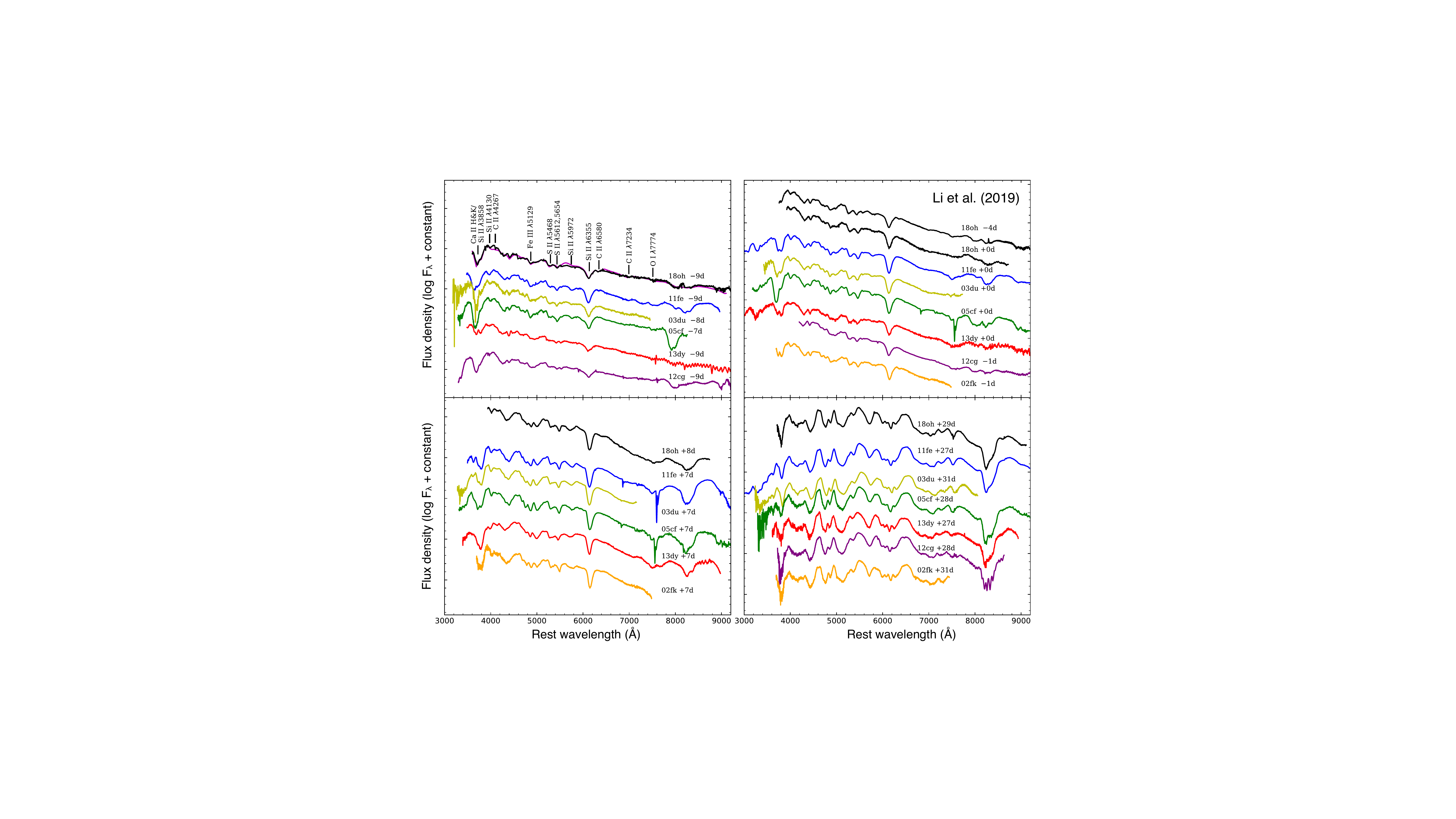}
\vskip -0.1in \caption{\textbf{Homogeneous optical spectra of \snia.} A
sample of \snia\ is shown at four epochs from $-$9 days to $+$1 month after
\textit{B}-maximum light. The main contributing features are shown via a model
fit (dark magenta) in the upper left panel. This figure is adapted from
ref.~\citen{Li2019}.
 \label{fig:li}}
\end{center}
\end{figure*}

\subsection{Spectral properties of \snia:}
The spectra of \sneia\ reveal the elements that are produced in the explosion,
their quantities, and their location within the SN ejecta (Figure
\ref{fig:li}). \snia\ spectra are dominated at early times and maximum light
by features from intermediate-mass elements such as calcium, magnesium,
silicon, and sulfur, with typical velocities measured from absorption minima
of 8000--15000 \kms\ around peak and decreasing with time as the photosphere
recedes. The earliest spectrum of a \snia\ is SN 2011fe at just over one day
past explosion\cite{Nugent2011}, and it was remarkably similar to
maximum-light spectra of \sneia, apart from the higher velocities. \snia\ show
a spectral sequence in which temperature, ionisation, and line ratios
correlate with peak luminosity\cite{Nugent95,Hachinger2008}.

After maximum light the spectra begin to be dominated by iron-group elements.
The ejecta expand with time, becoming optically thin by $\sim$150 days past
maximum light. \snia\ then enter the ``nebular''
phase\cite{Silverman13V,Graham2017} with spectra dominated by forbidden
emission lines of singly- and doubly-ionised iron (and other iron-group
elements such as cobalt and nickel; Figure \ref{fig:earlylate}). At $+$1000
days the spectrum of SN~2011fe showed a shift in ionisation to primarily
neutral iron\cite{Taubenberger2015,Fransson2015}.

\subsection{Performing cosmological measurements with \snia:} 

\begin{figure}[t]
\includegraphics[width=0.49\textwidth]{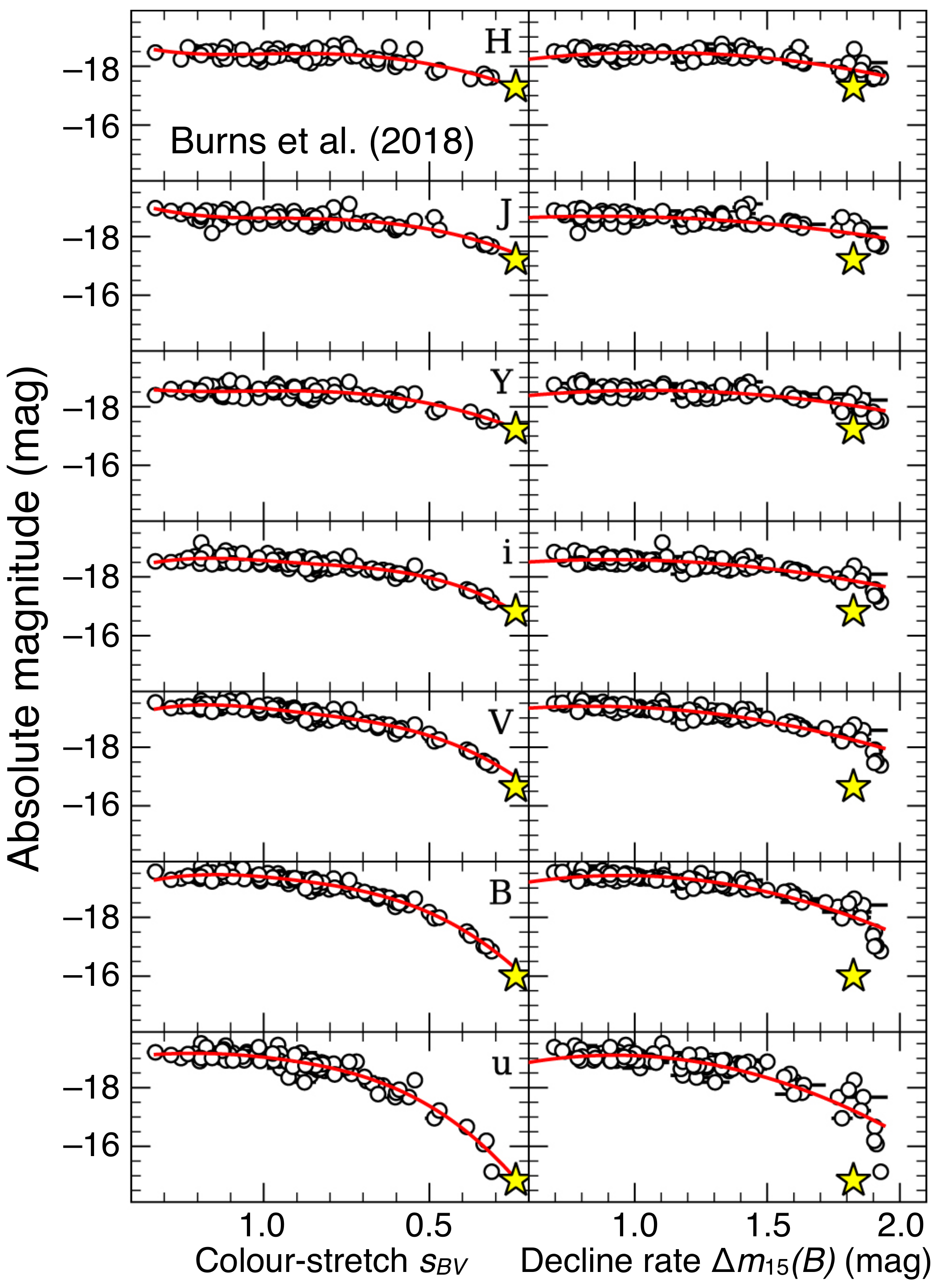}
\vskip -0.1in \caption{\textbf{Light curve shape standardisation of \snia.}
Modern versions of the Phillips relation\cite{Phillips93} from the Carnegie
Supernova Project\cite{Burns18}. The right panels use the original $\Delta
m_{15}(B)$ parameterisation, while the left panels use $s_{BV}$, the
light-curve colour-stretch\cite{Burns14}. Note the tight scatter around the
mean relations ($\sigma \lesssim$ 0.15 mag, except in \textit{u}) and the
flattening at longer wavelengths, showing that \snia\ are excellent
\emph{standard} (not just standardisable) candles in the near-infrared. The
yellow star is SN~2006mr, illustrating that the colour-stretch more
continuously parameterises fast-declining objects. This figure is adapted from
ref.~\citen{Burns18}.
\label{fig:burns}}
\end{figure}

As discussed above, \snia\ are best known as extragalactic distance indicators
and are essential in precision measurements of the cosmological parameters.
These measurements involve the use of empirical corrections to \snia\ light
curves to ``standardise'' their luminosities by correcting for the duration of
the light curve (light curve shape or ``stretch''), the optical colour at peak
brightness (``colour''), as well as a correction for the host galaxy
properties of the SN (see Section \ref{sec:environs}). The original
parameterisation of \snia\ light curves was based on the \textit{B} magnitude
decline in the 15 days after maximum light, $\Delta m_{15}(B)$; its
correlation with \snia\ luminosity is the Phillips relation\cite{Phillips93}.
Larger \snia\ samples have refined and extended this relation\cite{Hamuy96}
leading to the development of modern light-curve fitters to derive \snia\
distances\cite{Guy07,Jha07,Burns11}. Of particular recent note is the
expansion of the wavelength coverage of light-curve models to the near
infrared, where \snia\ appear to be more nearly
standard\cite{Krisciunas04,Barone-Nugent12,Dhawan18,Burns18} (rather than just
standardisable; Figure \ref{fig:burns}) and are less affected by dust
extinction.

\subsection{\snia\ progenitor systems and explosion mechanisms:} 
Understanding how and why stellar systems explode to produce \sneia\ is a
fundamental astrophysical question and relevant to more precise and accurate
\snia\ distances for future cosmological measurements. The research in this
area can be divided into two broad categories: studies looking for specific
\textit{signatures of the companion star} to the primary white dwarf and those
that try to unveil the \textit{explosion mechanism} that produces the
thermonuclear runaway that unbinds the star. The companion star of
\snia\ is thought to be either another degenerate white dwarf or a
non-degenerate star such as a main-sequence, giant, or helium star. As such
this question is often simplified to asking whether \snia\ arise from single-
or double-degenerate systems. However, many more questions remain as to how
the explosion begins and proceeds: what kind of material is accreted and how
quickly? at what mass does the primary white dwarf explode? does the explosion
start as a subsonic deflagration or a supersonic detonation? is the primary
white dwarf completely disrupted or is something left behind? what happens to
the companion star?

\subsection{Constraining companion stars:} 
Direct searches for companion stars to normal SN~Ia in data taken either
before\cite{Li11} or relatively soon (centuries)
after\cite{Schaefer2012,kerzendorf2013,kerzendorf2018,Ruiz-Lapuente18} the
explosion have not yielded any detections. Recently three hypervelocity stars
discovered in \emph{Gaia} data have been proposed as older surviving
white dwarf companions to \snia, arguing for a double-degenerate progenitor
system without complete disruption of the donor\cite{Shen18}.

\begin{figure*}
\hskip -0.35in 
\begin{center}
\includegraphics[width=0.75\textwidth]{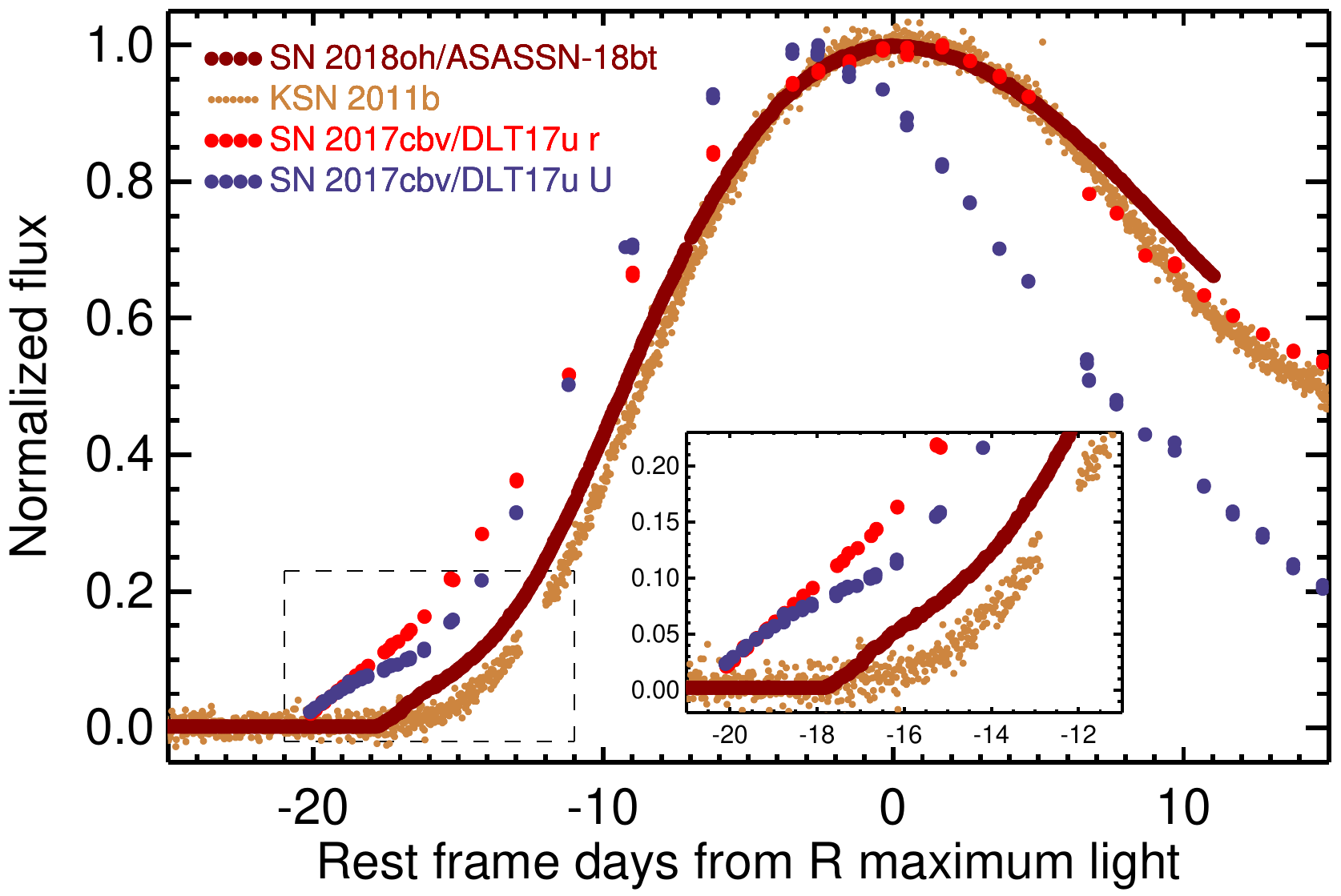}
\vskip -0.1in
\caption{\textbf{Early-time \snia\ light curves.} Exquisite \emph{Kepler}
light curves of SN~2018oh\cite{Dimitriadis2019,Shappee2019} and
KSN~2011b\cite{Olling15} and early-time ground-based LCOGT light curve of
SN~2017cbv\cite{Hosseinzadeh17} showing the diverse behaviour of
\snia\ in the days after explosion. The inset shows a zoom on the dashed box.
Deviations from a smooth early rise have been interpreted as shock interaction
with a non-degenerate companion, but they may alternatively result from
radioactivity in the outer SN ejecta.
\label{fig:kepler}}
\end{center}
\end{figure*}

\snia\ light curves within hours to days after explosion (Figure
\ref{fig:kepler}) can be used to search for potential shock interaction
between the SN ejecta and a companion star or other nearby material, and can
also probe properties of the exploding star such as the distribution of
nickel-56 and the ejecta density structure. Ground-based data are valuable if
there is fast-cadence monitoring of a SN field or if the SN is discovered
early\cite{Hosseinzadeh17}, but the most spectacular early supernova light
curves have been observed by the
\emph{Kepler}\cite{Olling15,Dimitriadis2019,Shappee2019} and now
\emph{TESS}\cite{Fausnaugh2019} spacecraft. Early-time ``bumps'' have clearly
been seen in SN~2017cbv and SN~2018oh (Figure \ref{fig:kepler}) but these are
not uniquely interpretable as companion shock
interaction\cite{Piro2013,Maeda18,Magee2018,Stritzinger18,Polin19}.
Observations have generally shown greater variety in \snia\ light curves at
early times compared to near or after maximum light\cite{Hayden10,Firth15}.

The presence of circumstellar material (CSM), more likely to arise in the
single-degenerate scenario, can be investigated using radio and X-ray
observations. The CSM is expected to be hydrogen-rich in the case of a
main-sequence companion star and helium-rich for a helium-star companion.
While X-ray and radio emission has been detected for some classes of
core-collapse supernovae, searches for X-ray and radio emission of \snia\ have
yielded only non-detections. The largest study to date of prompt radio
emission ($<$1 year after explosion) of 85 \snia\ resulted in non-detections
with upper limits on the pre-explosion mass loss rate, ruling out red giant
companions in $>$90\% of the sample\cite{chomiuk}. X-ray observations have
also placed constraining upper limits on the pre-explosion mass loss
rate\cite{russell12,horesh12,Margutti14}.

The CSM of \sneia\ can also be studied using the presence of narrow absorption
features of \NaiD\ and \Caii\ that are typically seen in the interstellar
medium but can also be present in \snia\ CSM. Blueshifted and time-varying
\NaiD\ features have been identified in a few \snia\ using high-resolution
spectroscopy\cite{Patat2007}; a recent example is SN~2013gh\cite{Ferretti16}.
Larger statistical samples have identified excess blueshifted \NaiD\ features
in \sneia, suggesting that there may be outflowing material (consistent with
CSM) present in $\sim$20\% of
\sneia\cite{Sternberg2011,Maguire2013,Phillips2013}. High-velocity ($>$15000
\kms) features of calcium (and sometimes silicon) are seen in early to
maximum-light spectra of $\sim$80\% \snia\cite{Mazzali2005} and are suggested
to indicate the presence of CSM or abundance enhancements in the SN
ejecta\cite{Zhao15}, but may also result from ionization effects in high
velocity material\cite{Blondin13}.

A handful of otherwise normal-looking \snia, like SN~2002ic\cite{Hamuy03} and
PTF~11kx\cite{Dilday2012}, have shown $H\alpha$ emission, taken as a sign of
interaction with hydrogen-rich material. These objects are generically
categorised as Ia-CSM\cite{Silverman13}. The strength and onset time of the
interaction can vary, but objects in this class are typically luminous,
slow-declining \snia\ in young environments. Late-time circumstellar
interaction is also proposed to explain the ultraviolet emission seen in
SN~2015cp nearly two years after the explosion\cite{Graham2019}.

Material from a hydrogen- or helium-rich companion star has been predicted to be
stripped (or ablated) during the explosion and result in the presence of
low-velocity hydrogen- or helium-rich material in the SN ejecta where it can be
energized by the radioactive decay and become visible. Searches for this
material have been made in many nearby \snia\ using late-time spectra, without
detection\cite{Mattila2005,Leonard2007,Lundqvist2013,Shappee2013,Lundqvist2015,Maguire2016,Graham2017,Shappee2018,Sand2018,Tucker2019,Sand19}.
This suggests that either the material is present and is not visible because it
is not located co-spatially with the radioactive material or that these objects
do not have hydrogen- or helium-rich companions.

A unique recent counterexample is ASASSN-18tb, a fast-declining \snia\ in an
early-type host galaxy, which showed nebular-phase $H\alpha$
emission\cite{Kollmeier2019}. However \emph{TESS} early-time observations of
ASASSN-18tb did not reveal a companion interaction signature and the nearly
constant $H\alpha$ flux may favour a circumstellar interaction power
source\cite{Vallely19} even though this supernova is quite different than
typical Ia-CSM.

\subsection{Understanding the explosion mechanism of \sneia:} 
The properties of the explosion mechanism are difficult to constrain because
of the complexities in the model predictions and the intrinsic variations in
observed \snia\ properties. Pre-maximum light spectra of \sneia\ can show the
presence of carbon (and oxygen) features that result from unburned material
from the exploding carbon-oxygen white dwarf. The most prominent optical
carbon feature is C II 6580 \AA\ and is seen in $>$40\% of \sneia\ with
spectra earlier than 10 days before maximum
light\cite{Parrent2011,Thomas2011,Blo12,Folatelli2012,Silverman2012,Maguire2014,Heringer2019}.
Oxygen can be studied via its 7773 \AA\ line, but there is difficulty
distinguishing between unburned oxygen and oxygen synthesised in the
explosion\cite{Zhao2016}. The velocity structure of the ejecta can be inferred
from spectral timeseries via abundance tomography and generally suggests
stratification in \snia\cite{Ashall14}, at odds with pure deflagration
(subsonic explosion) models\cite{Ma13}. The high ultraviolet opacity of
iron-group elements means that near-UV observations of \snia\ probe the outer
layers of the supernova and may provide another avenue to understanding the
pre-explosion composition and structure of the white dwarf and the explosion
mechanism\cite{Brown15,Foley16uv}. The abundances of stable iron-group
elements (requiring a neutron excess) can be inferred from late-time
spectra\cite{Maguire2018}; these are sensitive to the density of the burning
material and can help distinguish between Chandrasekhar mass (\MCh) and
sub-Chandra explosions\cite{Seitenzahl2017}.

\begin{figure*}
\hskip -0.35in 
\begin{center}
\includegraphics[width=\textwidth]{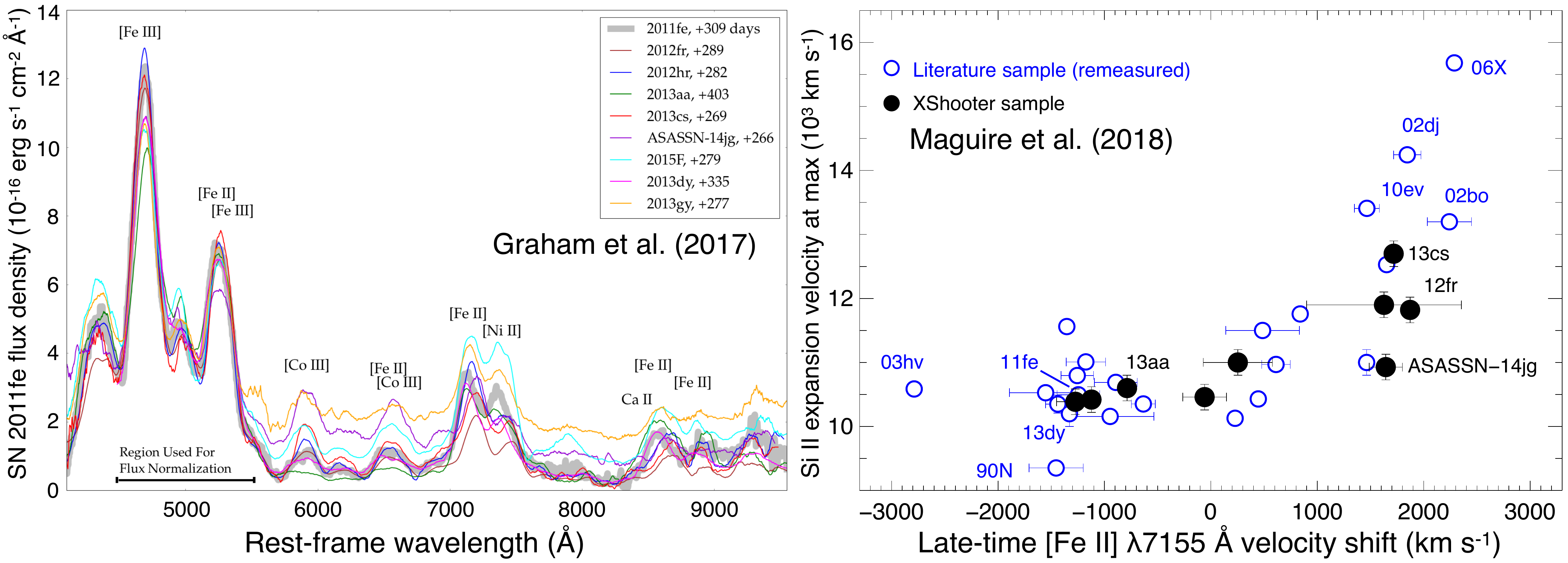}
\vskip -0.1in
\caption{\textbf{\snia\ nebular spectroscopy and line shifts.} Nebular spectra
of \snia\ (left panel) and velocity of the Si II absorption feature measured
at maximum light versus the velocity shift of the Fe II emission line in
late-time spectra of a sample of \snia\ (right panel). The error bars
displayed are 1-$\sigma$ uncertainties. A trend is seen where
\snia\ with redshifted Fe II features have higher maximum light Si II
velocities. The origin of this trend is debated but may be related to
asymmetries in the explosion mechanism and ejecta\cite{Maeda10}. This
figure is adapted from refs.~\citen{Graham2017} (left panel) and
\citen{Maguire2018} (right panel).
\label{fig:earlylate}}
\end{center}
\end{figure*}

Polarisation measurements of SN provide details on the geometry of the ejecta
and the extent of any asymmetries. Continuum polarisation is found to be small
in \snia, suggesting deviations of $<$10\% from spherical symmetry. However,
line polarisation has been found to be common in \snia, with significant
polarisation observed across the Si II and Ca II features that may suggest a
separate line-forming region or an asymmetric distribution for these elements
(at least at early times)\cite{Leonard2005,Wang2006,Patat2009}. Potential
asymmetries in the ejecta distribution (and hence the explosion) can also be
studied by looking at late-time spectra, where the outer layers have become
transparent and the core of the ejecta becomes visible. Shifts of up to
$\sim$3000 \kms\ (both to the blue and to the red) in iron and nickel
forbidden emission lines have been identified in spectra at $\sim$200 days
past maximum, indicating relatively large asymmetries in the inner Fe-rich
ejecta, and these seem to be correlated with early-time properties (Figure
\ref{fig:earlylate}), perhaps suggesting an orientation
effect\cite{Maeda10,Maguire2018}. Approximately 15\% of \snia\ show signatures
of double-peaked nebular emission lines separated by $\sim$5000
\kms, and the fraction rises for subluminous \snia\cite{Dong15,Vallely19b}; 
this may result from two explosion sites in a white dwarf collision model.

\section{\snia\ and their environments} 
\label{sec:environs}

\mbox{}

\sneia\ have been observed to occur in every type of galactic environment,
across galaxy types, stellar masses, metallicities and ages, from the lowest
mass dwarf galaxies to the most massive ellipticals. This simple observation
of ubiquity has significant implications for understanding the
\snia\ progenitor system and explosion physics: a \snia\ explosion must be
able to result from a progenitor with a wide range of stellar ages, from young
to very old systems. Detailed observations of \snia\ environments can provide
further clues.

\subsection{\snia\ rates: } The specific \snia\ rate (the \snia\ rate per unit
stellar mass) is significantly higher in star-forming later-type galaxies than
in early-type systems\cite{Vandenbergh90, Mannucci05}. Similar higher specific
rates are observed in bluer host galaxies compared to red host
galaxies\cite{Mannucci05}, lower mass galaxies compared to higher mass
galaxies\cite{Sullivan06,Brown19}, and in host galaxies with high specific
star-formation rates\cite{Sullivan06,Smith12}, i.e.~the star formation rate
(SFR) per unit stellar mass. The logical inference is that
\snia\ are more common in younger progenitor systems compared to older
progenitor systems, with a ``delay-time distribution'' (DTD\footnote{The
\snia\ DTD describes the \snia\ rate as a function of time following an
instantaneous burst of star formation. Thus it describes the likelihood of a
\snia\ explosion occurring as a function of the progenitor age.}) that
decreases sharply with progenitor age. More detailed analyses have shown that
these observations are a natural consequence of power-law
DTDs\cite{Kistler13,Graur15}.

These observations are consistent with the observed redshift evolution in the
volumetric \snia\ rate (the rate of \sneia\ per co-moving volume). The
volumetric \snia\ rate increases with increasing redshift, and by combining
volumetric rate measurements from different surveys across a range of
redshifts, several studies\cite{Maoz14,Graur14,Frohmaier19} have demonstrated
that this redshift evolution in the cosmic \snia\ rate is consistent with a
power-law DTD $\sim t^{-1}$, favouring double-degenerate progenitor systems.

\subsection{Environmental dependence of \snia\ properties: } The observed
variation in the rate of \sneia\ with the age of the progenitor stellar
population extends to some observed properties of \snia\ events. A key
observable affecting the utility of \sneia\ as cosmological probes is the
light-curve-width/luminosity relationship (Section~\ref{sec:snIa}): brighter
\sneia\ have light curves that evolve more slowly. It has been known for more
than twenty years that this light-curve width correlates with host galaxy
properties\cite{Branch96,Hamuy00,Sullivan06,Johansson13}, with brighter,
slower \sneia\ being hosted by younger, less massive, and more strongly
star-forming galaxies. This observation is, or should be, potentially
alarming: the fundamental standardising variable used in \snia\ cosmology
depends on the age of the \snia\ progenitor system. This implies that the
distribution of this parameter in \snia\ populations should evolve with
redshift, with a predicted shift to \sneia\ with brighter, slower light curves
at high redshift\cite{Howell07}. Such variations in \snia\ properties also
have implications for progenitor systems and explosion scenarios\cite{Shen17}.

\subsection{Impact on \snia\ distances and cosmology: }These relationships
between SN photometric properties and host galaxy properties demand that if
\snia\ are to be good standardisable candles over cosmic time, the calibrating
relationships between \snia\ luminosity and light-curve shape must be
invariant with progenitor age (or SN environment). In current samples, this
appears to be the case, at least to the level that it can currently be
measured. The relationships between luminosity and light-curve shape do not
show significant dependence on host galaxy properties.

More subtle trends in \snia\ properties as a function of environment have also
been detected: the standardised distance estimated from \sneia\ has a small
dependence on the properties of the SN host galaxies. This was originally
observed to occur as a function of the host galaxy stellar
mass\cite{Kelly10,Sullivan10,Lampeitl10}, with brighter \sneia\
(\textit{after} light-curve width and colour corrections) occurring in
higher-mass galaxies -- this is in the opposite sense to the far larger trend
between \textit{uncorrected} \snia\ luminosities and host properties. Galaxy
stellar mass is unlikely to be the fundamental variable or root cause of this
relationship -- galaxy stellar mass correlates with many other physical
quantities, such as metallicity (gas-phase and stellar), stellar age, and
galaxy dust content. The trend is also seen with many of these other global
host galaxy properties\cite{Childress13}, but global properties alone seem
unlikely to be able to resolve the physical cause of the variation, given the
covariance between the different variables. Selection effects, sample data
quality, and the choice of light curve standardisation may also play a
role\cite{Scolnic18,Brout19}.

\begin{figure*}
\hskip -0.35in 
\begin{center}
\includegraphics[width=0.975\textwidth]{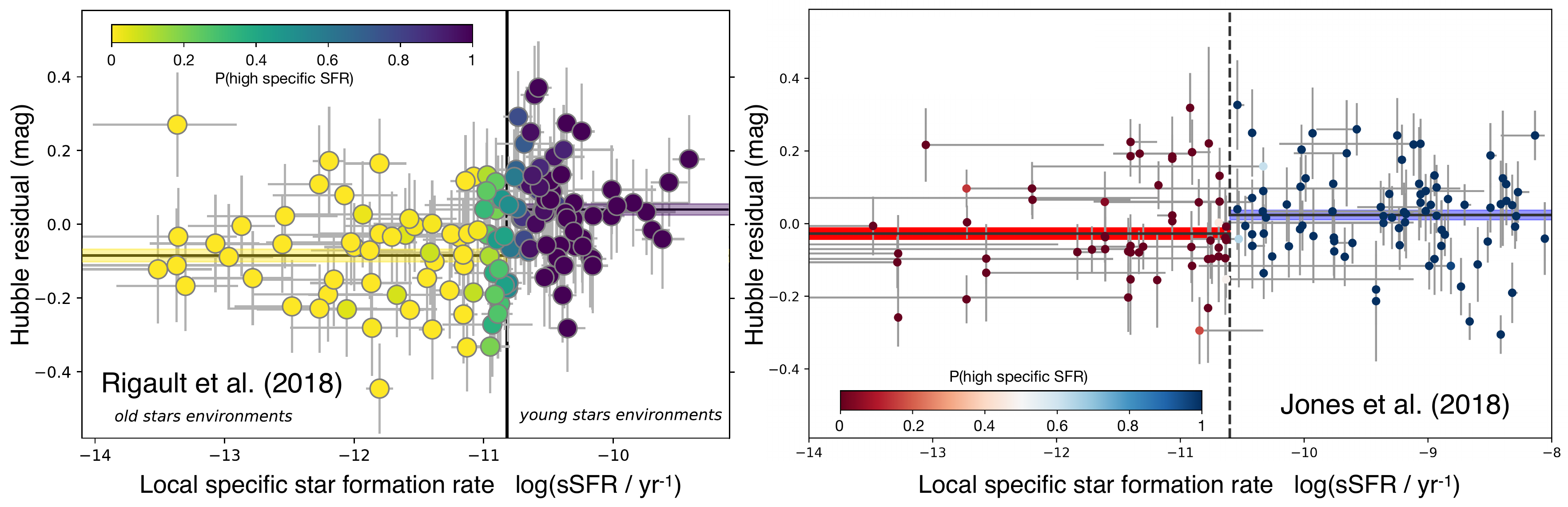}
\vskip -0.1in
\caption{\textbf{The effect of environment on \snia\
standardisation.} \snia\ Hubble diagram residuals as a function of the local
specific star-formation rate (sSFR), the star formation rate per unit stellar
mass. Left: The sSFR is measured using H$\alpha$ emission measured in a 1\,kpc
projected average radius; Right: The sSFR is measured using SED fitting to
broad-band optical data in 1.5\,kpc projected apertures. The colours show the
probability for a SN to have a younger environment. The two horizontal bands
give the weighted average of the Hubble residuals per sSFR group. The width of
each band represents the corresponding error on the mean, and their offset
illustrates the Hubble residual offset between the two sSFR groups. The error
bars displayed are 1-$\sigma$ uncertainties. This figure is adapted from
refs.~\citen{Rigault19} (left panel) and \citen{Jones18b} (right panel).
\label{fig:environment}}
\end{center}
\end{figure*}

Measurements of \emph{local} galaxy properties at the SN position hold promise
to clarify the picture (Figure~\ref{fig:environment}). \snia\ Hubble residuals
show a correlation with local specific star formation
rate\cite{Rigault13,Rigault19,Jones18b}, measured either with nearby
nebular $H\alpha$ emission, rest-frame near-UV flux, or galaxy SED fitting to
optical colours, though there remains some disagreement as to whether
corrections based on these local measures should be used in preference to
those based on global host properties. Most of this work has focused on
\sneia\ in the local universe, where such studies are simpler to perform as
the \snia\ host galaxies are better resolved. However, a similar trend is also
seen at moderate redshift using local photometry\cite{Rose19} and in
high-redshift samples using rest-frame UV photometry measured in 3\,kpc
apertures from deep imaging stacks\cite{Roman18}. \snia\ in locally
star-forming regions may also have a lower dispersion than the
rest\cite{Kelly15,Rigault19}.

The interpretation of these results is unclear. There is a prediction
from empirical galaxy models that low-mass galaxies should be expected to
contain a more homogeneous population of young \snia\ progenitors across all
redshift ranges\cite{Childress14}. Assuming these lower-mass galaxies are also
more strongly star-forming, this is consistent with observations that
\sneia\ in star-forming galaxies present a more homogeneous population.
Selecting these events in cosmological studies may therefore provide access to
a \snia\ sample with a narrow range in progenitor ages, therefore removing the
challenge of using corrections in cosmological analyses based on host
environment, and perhaps reducing potential astrophysical systematic effects
when using \sneia\ in cosmology.

\section{The Thermonuclear Supernova Zoo}
\label{sec:zoo}

\begin{figure*}
\hskip -0.35in 
\begin{center}
\includegraphics[width=0.9\textwidth]{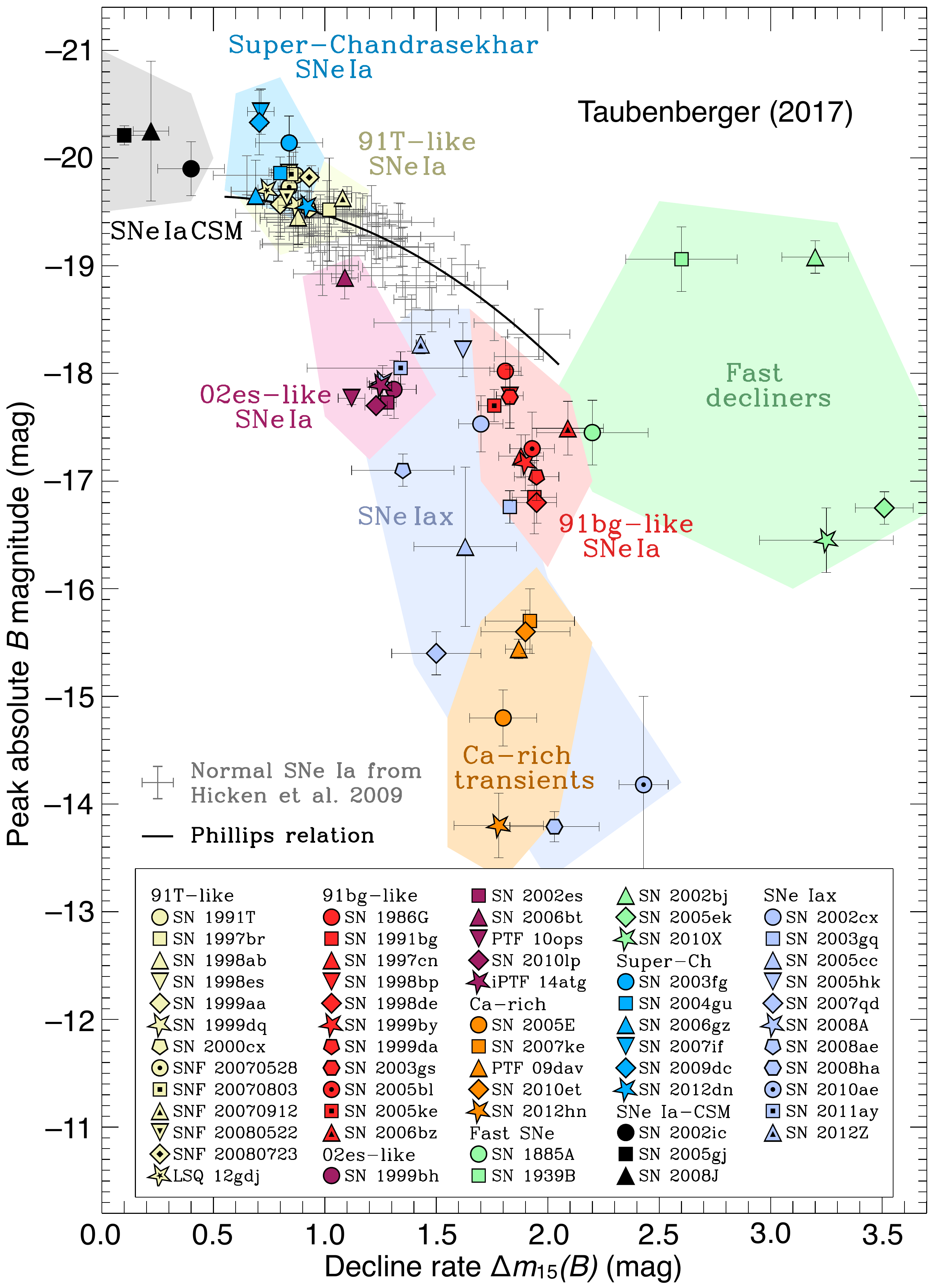}
\vskip -0.1in
\caption{\textbf{The thermonuclear supernova zoo.} Luminosity versus
light-curve decline rate of normal \snia, extreme \snia, and the wide variety
of peculiar white dwarf supernovae. The error bars displayed are 1-$\sigma$
uncertainties. This figure is adapted from ref.~\citen{Taubenberger17}.
\label{fig:taubenberger}}
\end{center}
\end{figure*}

\mbox{}

The Phillips relation\cite{Phillips93} defines a one-parameter family of
\snia, seen also as a spectral sequence\cite{Nugent95}. The slow-declining,
hot, luminous end is marked by 91T-like or 99aa-like objects, showing
prominent Fe III features with weak Si II in maximum-light spectra. 99aa-like
\snia\ also show strong Ca II absorption that is much weaker in 91T-like
objects\cite{Silverman12,Taubenberger17}. The fast-declining, cool,
subluminous 91bg-like \snia\ are most often found in old stellar
populations\cite{Panther19} (i.e., passive host galaxies) and are sometimes
claimed to be a separate population from more normal \snia, with few
``transitional'' objects in between\cite{Srivastav17,Cartier2017}. However,
this may only be pointing to a shortcoming of the $\Delta m_{15}$
parameterisation\cite{Phillips93}; using colour\cite{Garnavich04} or
colour-stretch\cite{Burns14,Burns18} suggests a more continuous distribution
with other \snia\ (Figure \ref{fig:burns}). Off the Phillips relation is the
realm of ``peculiar'' thermonuclear supernovae. The same
luminosity/decline-rate parameter space has been used to distinguish these
objects\cite{Taubenberger17} (Figure \ref{fig:taubenberger}).

\begin{figure*}
\hskip -0.35in 
\begin{center}
\includegraphics[width=0.92\textwidth]{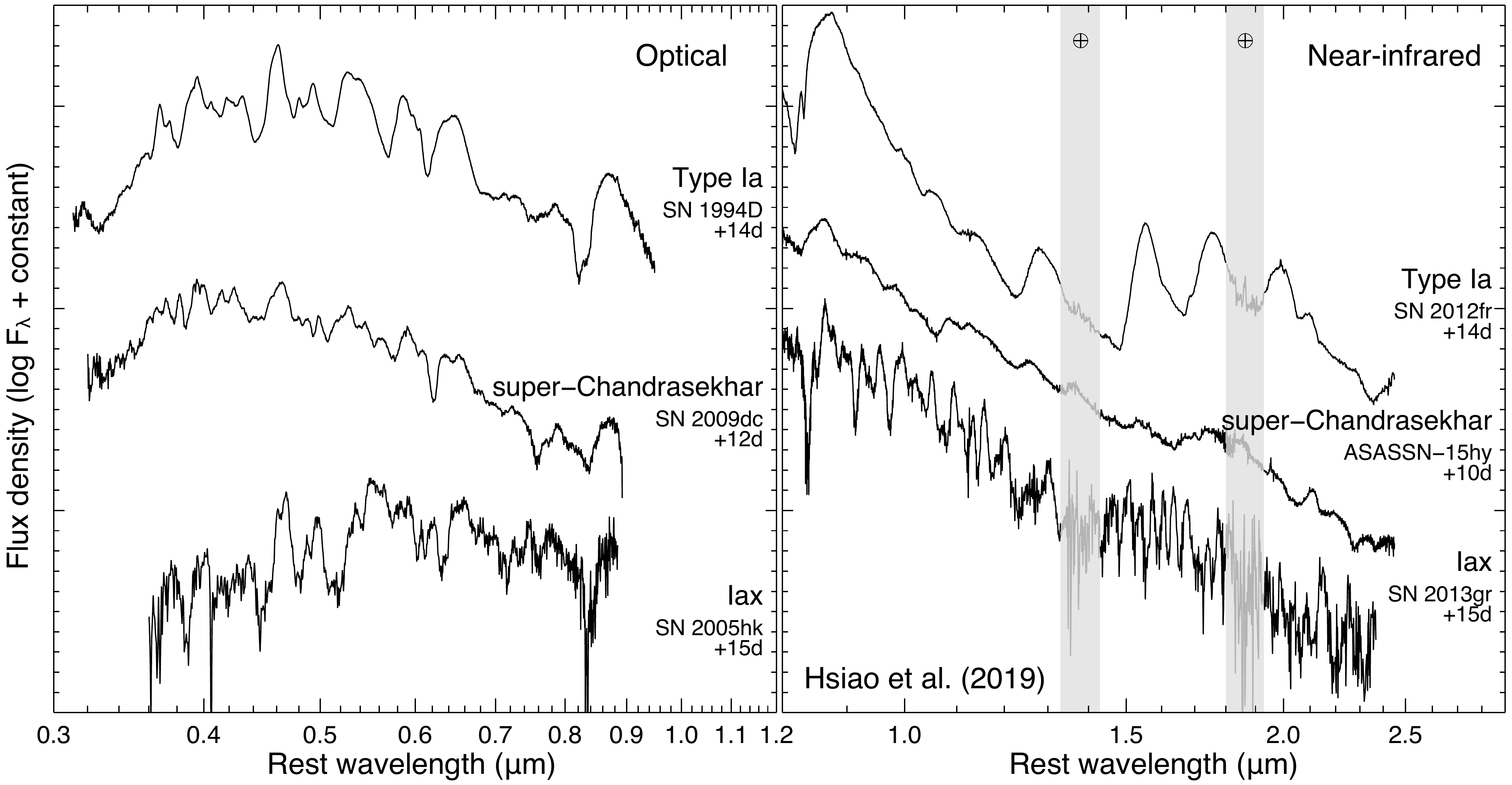}
\vskip -0.15in
\caption{\textbf{Optical and near-infrared spectroscopy of thermonuclear
supernovae.} Near-infrared spectroscopy (right) illustrates the differences
between subclasses of thermonuclear supernovae more clearly than optical
spectra (left). This figure is adapted from ref.~\citen{2019PASP..131a4002H}.
 \label{fig:hsiao}}
\end{center}
\end{figure*}

Several groups of white dwarf supernovae show low ejecta velocities, below the
typical 10,000 \kms\ Si II velocity seen in normal \sneia\ around maximum
light. The most numerous of these are type Iax supernovae\cite{Foley13,Jha17}
(SN Iax) with SN~2002cx as the prototype\cite{Li03,Jha06}. Though typically
found in star-forming environments, SN~Iax are thought to be white-dwarf
supernovae because of their spectral similarity to normal \snia\ at early
times, including the dominance of iron-group elements (like the Co II infrared
lines\cite{Stritzinger15,2016MNRAS.459.1018T}; Figure \ref{fig:hsiao}) as well
as their radioactively powered light curves\cite{McCully14}. Near maximum
light SN~Iax have low photospheric velocities (from 7000 down to 2000 \kms)
and typically low luminosity ($-19 \lesssim M_V \lesssim -13$) compared to
normal \snia, and show more overall diversity\cite{Jha17}. At late times
SN~Iax differ from all other supernovae in never becoming fully ``nebular'' in
their spectra, with the marked presence of low velocity ($<$ 2000 \kms)
permitted iron lines. The SN~Iax 2012Z is unique among all thermonuclear
supernovae because of the detection of its luminous progenitor system in
prediscovery \emph{Hubble Space Telescope} images\cite{McCully14_12Z},
interpreted as a helium-star donor to the exploding white dwarf. A leading
model for SN~Iax is the pure-deflagration explosion of a Chandrasekhar-mass
white dwarf in a helium-accreting single-degenerate system\cite{Jha17}. It is
possible this explosion does not completely disrupt the white dwarf, leaving a
bound remnant\cite{Kromer13}. Surviving examples of such incomplete explosions
may have been discovered in our Galaxy as fast-moving white dwarfs with
unusual abundances\cite{Vennes17,Raddi19}.

Perhaps related to the class of SN~Iax are
02es-like\cite{White15,Taubenberger17} SN which also have low luminosity and
low ejecta velocity, but ``cool'' spectra compared to the ``hot'' SN~Iax.
Those properties and the preference for 02es-likes for old stellar
environments are similar to 91bg-like \snia, but 02es-likes do not have fast
light-curves. The 02es-like iPTF~14atg showed evidence of an early-time ``blue
bump'' that may be consistent with shock interaction of the supernova ejecta
with a companion star\cite{Cao15}. SN~2010lp was an 02es-like object that
showed nebular oxygen emission\cite{Taubenberger13b} and a similar feature was
observed in iPTF~14atg\cite{Kromer16}; such emission has not been seen in any
other thermonuclear SN.

``Super-Chandrasekhar'' \snia\ have optical spectra similar to normal
\snia, though with strong carbon features near maximum light and more distinct
near-infrared spectra (Figure \ref{fig:hsiao}). They have relatively high
luminosity, slow light curves, and low ejecta velocities. Taken together these
properties imply a high ejecta mass, exceeding \MCh, thus explaining the
``super-Chandra''
moniker\cite{Howell06,Taubenberger11,Taubenberger13,Taubenberger17,Scalzo19}.
An observationally-based name for this class may be preferable to a
model-dependent one\cite{Chen19}, but these objects and their progenitors
regardless raise fundamental astrophysics questions. One clue to their nature
may come from a preference for low-metallicity environments\cite{Khan11}.

Another class of objects for which environments may be the key to
understanding is ``calcium-rich'' supernovae. Spectroscopically similar to
type-Ib supernovae, with prominent helium features at maximum light, these
low-luminosity explosions occur far from any star formation, and indeed often
far from their host galaxies with almost no underlying stellar
light\cite{Perets11,Kasliwal12,Lyman13,Lunnan17}. In their nebular spectra
these objects are dominated by strong [Ca II] emission, giving them their
name. A proposed origin is a long delay-time thermonuclear explosion of a
white dwarf in a binary system that was dynamically ejected from its
host\cite{Perets10,Foley15ca}, but multiple progenitor scenarios may be
required\cite{Milisavljevic17,De18}.

The thermonuclear supernova zoo also contains unusual objects which have not
(yet) been easily grouped into classes, e.g., SN~2000cx and its twin
SN~2013bh\cite{Li01,Candia03,Silverman13b}; ``fast and faint'' objects like
SN~2005ek\cite{Drout13}, PTF09dav\cite{Sullivan11}, SN~2010X\cite{Kasliwal10},
and even SN~1885A\cite{Fesen17}; the slow and faint PTF10ops\cite{Maguire11};
and the fast and not-so-faint SN~1939B\cite{Perets11b}, and
SN~2002bj\cite{Poznanski10}. Unique objects continue to be discovered, like
the high-velocity SN~2019ein\cite{Burke19}. A number of peculiar thermonuclear
transients have shown evidence for detonation of a helium
shell\cite{Jiang17,De19} and may imply a diversity in total mass and shell
mass for exploding white dwarfs\cite{Polin19}. Though the zoo is stocked with
a broad variety of thermonuclear SN, volumetric rates of these species ``in
the wild'' can vary widely. The luminous peculiar objects, like super-Chandra
or Ia-CSM, are intrinsically rare and not more than a few percent of the
\snia\ rate\cite{Gal-Yam2017}. Subluminous peculiar objects are more
common\cite{Foley13,Miller17,Frohmaier18}, but it is nonetheless likely that
normal \snia\ that lie on the Phillips relation still comprise the most
numerous class of thermonuclear supernovae. This remarkable fact needs an
explanation.

In this review we have tried to highlight recent advances in our observational
understanding of thermonuclear supernovae. Even with this limited aim our
review is incomplete, and moreover, we have not been able to sufficiently
discuss important progress from theory and computation on models of white
dwarf supernova progenitor systems and explosions. These shortcomings testify
to the vibrancy of the field. We should make special note that the bulk of the
observational progress described here is predicated on the increasing number
of bright or nearby supernovae (and their earlier discovery) from a number of
surveys like ASASSN\cite{ASASSN17}, ATLAS\cite{ATLAS18}, CRTS\cite{CRTS09},
DLT40\cite{DLT40}, \emph{Gaia}\cite{Gaia13}, LOSS\cite{LOSS01},
LSQ\cite{LSQ15}, MASTER\cite{MASTER10}, OGLE\cite{OGLE14},
Pan-STARRS\cite{PSST15}, PTF/iPTF\cite{PTF09}, PTSS\cite{PTSS16},
ZTF\cite{ZTF19}, among others, and the continued work of amateur astronomers.
Upcoming surveys like LSST\cite{LSST19} will provide large samples of more
distant supernovae, including rare objects, and will allow nearby supernovae
to be systematically observed to late times as they fade. Such samples will
help develop a deeper physical understanding of thermonuclear supernova
progenitors and explosions and can be used to improve distances from \snia\ and
measurements of cosmological parameters.

\bibliography{thermonuclear}

\clearpage 

\begin{addendum}

\item[Correspondence] 
Correspondence and requests for materials should be addressed to
S.W.J.~(saurabh@physics.rutgers.edu).

\item[Acknowledgements]
We thank Chris Burns, R\'{e}gis Cartier, Melissa Graham, Eric Hsiao, David
Jones, Wenxiong Li, Mikael Rigault, and Stefan Taubenberger for providing
figures used here. We are grateful to Ryan Foley, Avishay Gal-Yam, Peter
Nugent, Ken Shen, and the anonymous referee for comments and helpful
suggestions. We also thank Fiona Panther for alerting us to the Ph.D.~thesis
of Titus Pankey, Jr.~as the genesis of a radioactive nickel-56 power source
for the luminosity of supernovae. Support for this review was provided in part
by US NSF award AST-1615455 (S.W.J.), H2020/ERC grant 758638 (K.M), and FP7
EU/ERC grant 615929 (M.S.).

\item[Author Contributions] 
S.W.J.~wrote the introduction and thermonuclear zoo section of the article.
K.M.~wrote the section on \snia\ observations, and M.S.~wrote the section on
\snia\ environments. All authors discussed and edited the text.

\item[Competing Interests] 
The authors declare that they have no competing financial interests.

\end{addendum}

\end{document}